\documentclass[12pt,a4paper]{article}

\usepackage{amsmath}
\usepackage{amssymb}
\usepackage{hyperref}
\usepackage{dsfont}
\usepackage{hyperref}
\usepackage{graphics}
\usepackage{graphicx}
\usepackage{epsfig}

\let\oldappendix=\appendix
\let\oldsection=\section
\renewcommand{\appendix}{\oldappendix%
\def\theequation{\Alph{section}.\arabic{equation}}%
\renewcommand{\section}{\setcounter{equation}{0}\oldsection}}

\newcommand{\beq}{\begin{equation}}
\newcommand{\eeq}{\end{equation}}
\newcommand{\beqa}{\begin{eqnarray}}
\newcommand{\eeqa}{\end{eqnarray}}
\newcommand{\no}{\nonumber}
\newcommand{\q}{\quad}
\newcommand{\qq}{\qquad}

\newcommand{\sfrac}[2]{{\textstyle\frac{#1}{#2}}}

\newcommand{\newop}[2]{\def#1{\mathop{\mathrm{#2}}\nolimits}}
\newop{\artanh}{artanh}

\newcommand{\mcirc}[2][2]{{\overset{\circ}{m}\vphantom{m}_{#2}^{#1}}}

\newcommand{\decay}{f}

\begin{document}

\hfill 

\hfill 

\bigskip\bigskip

\begin{center}

{{\Large\bf  The number of colors\\[0.3cm]
            in the decays $\mbox{\boldmath$\pi^0, \eta, \eta' \rightarrow \gamma \gamma$}$ }}

\end{center}

\vspace{.4in}

\begin{center}
{\large B. Borasoy\footnote{email: borasoy@ph.tum.de}}

\bigskip

\bigskip

\href{http://www.ph.tum.de/}{Physik Department}\\
\href{http://www.tum.de/}{Technische Universit{\"a}t M{\"u}nchen}\\
D-85747 Garching, Germany \\

\vspace{.2in}

\end{center}

\vspace{.7in}

\thispagestyle{empty} 

\begin{abstract}
The decays $\pi^0, \eta, \eta' \rightarrow \gamma \gamma$
are investigated up to next-to-next-to leading order in the framework
of the combined $1/N_c$ and chiral expansions. Without mixing of the
pseudoscalar mesons the $N_c$ independence of the $\pi^0$ and $\eta$ decay 
amplitudes is shown to persist at the one-loop level, although the contribution
of the Wess-Zumino-Witten term to the pertinent vertices is not canceled by the $N_c$ dependent part
of a Goldstone-Wilczek term.
The decay amplitude of the singlet field, on the other hand, depends strongly
on $N_c$ and yields under the inclusion of mixing also a strong $N_c$ dependence
for the $\eta$ decay. Both the $\eta$ and $\eta'$ decay are suited to confirm the
number of colors to be $N_c=3$.

\end{abstract}\bigskip

\begin{center}
\begin{tabular}{ll}
\textbf{PACS:}&12.39.Fe, 13.40.Hq\\[6pt]
\textbf{Keywords:}& Electromagnetic decays, chiral Lagrangians, large $N_c$.
\end{tabular}
\end{center}

% 11.30.Rd Chiral symmetries
% 12.39.Fe Chiral Lagrangians
% 12.40.Vv Vector-meson dominance
% 13.20.Cz Decays of pi mesons
% 13.40.Hq Electromagnetic decays

\vfill

%%%%%%%%%%%%%%%%%%%%%%%%%%%%%%%%%%%%%%%%%%%%%%%%%%%%%%%%%%%%%%%%%%%%%%%%%%%%%%
%%%%%%%%%%%%%%%%%%%%%%%%%%%%%%%%%%%%%%%%%%%%%%%%%%%%%%%%%%%%%%%%%%%%%%%%%%%%%%
\section{Introduction}\label{sec:intro}

Experimental evidence suggests that we live in a world with three colors.
At high energies the Drell ratio for $e^+ e^-$ annihilation supports $N_c =3$,
while in the low-energy regime below 1 GeV the anomalous decay of the $\pi^0$
into two photons , $\pi^0 \rightarrow \gamma \gamma$, is presented as a 
textbook example to confirm the number of colors, see {\it e.g.} \cite{DGH}.
The quark charges are assumed to be independent of $N_c$ yielding at tree level
a width $\Gamma_{\pi^0 \rightarrow \gamma \gamma}$ proportional to $N_c^2$,
thus being quite sensitive to the number of colors. However, it was shown recently
in \cite{BW} that the cancellation of triangle anomalies in the standard model 
with an arbitrary number of colors leads to $N_c$ dependent values of the quark 
charges. For three light flavors $(u,d,s)$ $N_c$ enters as a quantized prefactor
of the Wess-Zumino-Witten (WZW) term \cite{WZ,W}, but the vertex with one pion
and two photons is completely canceled by the $N_c$ dependent part of a
Goldstone-Wilczek term \cite{BW, GW}. A similar cancellation also occurs for the
decay $\eta \rightarrow \gamma \gamma$, if one neglects $\eta$-$\eta'$ mixing.

On the other hand, the quark triangle diagram of the microscopic theory describing the decay 
$\eta_0 \rightarrow \gamma \gamma$ of the flavor singlet is $N_c$ dependent
and hence due to $\eta$-$\eta'$ mixing the decay width
of the $\eta$ will also pick up an $N_c$ dependent portion.
If one works with different up- and down-quark masses, $m_u \ne m_d$, the
$\pi^0$ will also undergo mixing with the $\eta$-$\eta'$ system and its decay
width into two photons will have an---albeit small---$N_c$ dependent piece.

Loop diagrams which have not been discussed in \cite{BW} can be another source
of $N_c$-dependence for the two-photon decays. As we will see, the vertices
of the one-loop diagrams contain indeed an $N_c$ dependent piece that does not
cancel out in the sum of the WZW and Goldstone-Wilczek terms.

In order to investigate systematically the effects of mixing and loops
for the two-photon decays of $\pi^0, \eta$ and $\eta'$, we include
the $\eta'$ explicitly within the combined framework of chiral perturbation theory (ChPT)
and the $1/N_c$ expansion, so-called large $N_c$ ChPT \cite{KL1, KL2, H-S}.\footnote{
For an alternative approach to include the $\eta'$ without employing large $N_c$
counting rules, see, {\it e.g.}, \cite{etap}.}
In this theory, the $\eta_0$ is combined with the octet of pseudoscalar mesons
$(\pi, K, \eta_8)$, since in the large $N_c$ limit the axial $U(1)$ anomaly vanishes
and the $\eta'$ converts into a Goldstone boson.
In the present work, we evaluate the decay amplitudes of $\pi^0, \eta$ and $\eta'$
up to next-to-next-to-leading order at which loops start contributing
in large $N_c$ ChPT.

The paper is organized as follows. In the next section, we discuss the WZW term
under the inclusion of the $\eta'$. We will see that it can be decomposed into
the conventional $SU(3)$ WZW Lagrangian, the Goldstone-Wilczek term and counter terms
of unnatural parity. In Sec.~3 the calculation for the decays up to
next-to-next-to-leading order is presented. Numerical results and the importance
of $N_c$ dependent contributions are discussed in Sec.~4.
Sec.~5 contains our conclusions and the scaling behavior of the coupling constants
under changes of the QCD running scale is presented in App.~\ref{app:scal}.

%%%%%%%%%%%%%%%%%%%%%%%%%%%%%%%%%%%%%%%%%%%%%%%%%%%%%%%%%%%%%%%%%%%%%%%%%%%%%%
\section{Wess-Zumino-Witten term}\label{sec:wzw}

In this section we will first briefly outline the method of extending
the $SU(3)_R \times SU(3)_L$ chiral rotations of the effective Lagrangian
in conventional ChPT to $U(3)_R \times U(3)_L$  in a more generalized framework 
including the $\eta'$ \cite{KL1, H-S}. Within this approach the topological
charge operator coupled to an external field $\theta$ is added to the QCD Lagrangian
\beq
{\cal L} = {\cal L}_{\scriptscriptstyle{QCD}} - \frac{g^2}{16 \pi^2} \theta (x) \mbox{tr}_c
            ( G_{\mu \nu} \tilde{G}^{\mu \nu})
\eeq
with $G_{\mu \nu}$ the gluonic field strength tensor, $\tilde{G}_{\mu \nu}= 
\epsilon_{\mu \nu \alpha \beta} G^{\alpha \beta}$ its dual counterpart, and
$\mbox{tr}_c$ is the trace over the color indices. Under $U(1)_R \times U(1)_L$
the axial $U(1)$ anomaly adds a term $- \frac{g^2}{16 \pi^2} N_f (\alpha_R - \alpha_L) \mbox{tr}_c
( G_{\mu \nu} \tilde{G}^{\mu \nu})$ to the QCD Lagrangian, with $N_f$ being the number
of different quark flavors and $\alpha_{R/L}$ the angle of the $U(1)_{R/L}$ rotation.
The vacuum angle $\theta (x)$ is in this context treated as an external pseudoscalar
source that transforms under axial $U(1)$ rotations as
\beq
\theta (x) \rightarrow \theta' (x) = \theta (x) + i \ln \det R - \ln \det L
\eeq
with $R,L \in U(1)$, so that the term in the anomaly proportional to the topological
charge operator of the gluons is compensated by the shift in the $\theta$ field.
There are, however, further axial anomalies which are accounted for within the effective
theory by the WZW term.
The dynamical variables of the effective theory are the  
pseudoscalar mesons $(\pi, K, \eta_8, \eta_0)$  that live in the coset space
$U(3)_R \times U(3)_L/ U(3)_V = U(3)$. They are most conveniently collected
in a unitary matrix $U(x) \in U(3)$ with a phase given by
\beq
\det U (x) = e^{i \psi (x)} .
\eeq
The field $\psi$ describes the singlet field $\eta_0$ and 
is the extension from the standard framework where
the effective field is an element of $SU(3)$.
Under chiral rotations, the effective field $U(x)$ transfroms as
\beq
U'(x) = R(x) U(x) L^\dagger (x) ,
\eeq
so that its phase changes by
\beq
\psi'(x) = \psi (x) - i \ln \det R + \ln \det L .
\eeq
Hence, the combination $\bar{\psi} = \psi + \theta$ remains invariant in the effective
theory under chiral $U(3)_R \times U(3)_L$ rotations.
The covariant derivative of $U$ involves left- and right-handed sources
\beq
D_{\mu} U  =  \partial_{\mu} U - i r_{\mu} U  + i U l{\mu}  
\eeq
with $r_{\mu} = v_{\mu} + a_{\mu}$ and $l_{\mu} = v_{\mu} - a_{\mu}$.
In terms of these building blocks the WZW effective action is given by \cite{WZ, W}
\beqa
S_{\scriptscriptstyle{WZW}} (U, v, a) &=& S_{\scriptscriptstyle{WZW}} (U) +
S_{\scriptscriptstyle{WZW}} (v,a)  -  \frac{i N_c}{48 \pi^2} \int \; 
\langle U \, l^3  U^\dagger  r + \sfrac{1}{4} U \, l \, U^\dagger r 
\, U \, l \, U^\dagger  r \no \\ 
&& + i U \, dl \, l \, U^\dagger r  + i dr \, U \, l \, U^\dagger r 
- i \Sigma_{\scriptscriptstyle{L}} l \, U^\dagger r \, U \, l 
   + \Sigma_{\scriptscriptstyle{L}}  U^\dagger dr \, U \, l 
   - \Sigma_{\scriptscriptstyle{L}}^2 U^\dagger r \, U \, l
  \no \\ 
&&+ \Sigma_{\scriptscriptstyle{L}} l \, dl  + \Sigma_{\scriptscriptstyle{L}} dl \, l 
  - i \Sigma_{\scriptscriptstyle{L}} l^3 
  + \sfrac{1}{2} \Sigma_{\scriptscriptstyle{L}} l \, \Sigma_{\scriptscriptstyle{L}} l
  - i \Sigma_{\scriptscriptstyle{L}}^3 l \rangle -  (R \leftrightarrow L) ,
\eeqa
where $\Sigma_{\scriptscriptstyle{L}} = U^\dagger dU $ 
and we adopted the differential form notation of \cite{KL1},
\beq
v = dx^\mu v_\mu , \qq a = dx^\mu a_\mu , \qq r = v+a , \qq l=v-a ,\qq d = dx^\mu \partial_\mu .
\eeq
with the Grassmann variables $dx^\mu$ which yield the volume element $dx^\mu
dx^\nu dx^\alpha dx^\beta = \epsilon^{\mu \nu \alpha \beta} d^4x$.
The brackets $\langle \ldots \rangle$ denote the trace in flavor space and
the operation $(R \leftrightarrow L)$ indicates the interchange of $r$ with $l$ as
well as of $U$ with $U^\dagger$, so that, {\it e.g.}, $\Sigma_{\scriptscriptstyle{L}}$ is
replaced by $\Sigma_{\scriptscriptstyle{R}} = U dU^\dagger $.

In order to extract the $SU(3)$ version of the WZW term, it is convenient to introduce
the notation
\beq
U = e^{\frac{i}{3} \bar{\psi}} \bar{U} , \qquad \det \bar{U} = e^{-i \theta} .
\eeq
As the field $\bar{\psi} = \psi + \theta$ is gauge invariant, $\bar{U}$ transforms in the same
manner as $U$ under chiral rotations and its covariant derivative is defined as
\beqa
D_{\mu} \bar{U}  &=&  \partial_{\mu} U - i (v_{\mu} + \bar{a}_\mu ) U + i U ( v_{\mu} -\bar{a}_\mu ), \no \\ 
\bar{a}_\mu &=& a_\mu - \sfrac{1}{3} \langle a_\mu  \rangle - \sfrac{1}{6} \partial_\mu \theta
   =  a_\mu  -  \sfrac{1}{6} D_\mu \theta   .
\eeqa
In \cite{KL1} it has been shown that the WZW term can be decomposed as
\beq  \label{eq:wzwsu3}
S_{\scriptscriptstyle{WZW}} (U, v, a) = S_{\scriptscriptstyle{WZW}} (\bar{U}, v, \bar{a})
   + \int B
\eeq
with
\beqa
B   &=& - \frac{N_c}{144 \pi^2} \Big( \bar{\psi}  \langle i F_{\bar{r}} D \bar{U} D \bar{U}^\dagger 
    + i F_{\bar{l}} D \bar{U}^\dagger D \bar{U} + 2 F_{\bar{r}} \bar{U} F_{\bar{l}} \bar{U}^\dagger
   + 2  F_{\bar{r}}^2 + 2 F_{\bar{l}}^2 \rangle  \no \\
  && \qquad \qquad 
      + \sfrac{1}{6} \bar{\psi}  \langle F_{\bar{r}} -F_{\bar{l}} \rangle 
      \langle F_{\bar{r}} -F_{\bar{l}} \rangle 
      - i D \theta \langle  F_{\bar{r}} D \bar{U} \bar{U}^\dagger  
      - F_{\bar{l}} \bar{U}^\dagger D \bar{U}  \rangle \Big)      
\eeqa
and
\beq
F_{\bar{r}} = d{\bar{r}} -i{\bar{r}}^2, \qquad F_{\bar{l}} = d{\bar{l}} -i{\bar{l}}^2 .
\eeq
The quantities $\bar{r},\bar{l}$ are the QCD renormalization group invariant parts of the
left- and right-handed gauge fields $r= \bar{r} + \frac{1}{6} D \theta, l= \bar{l} - 
\frac{1}{6} D \theta$ with $D \theta = d \theta + 2 \langle a \rangle$.
The left- and right-hand side of Eq.~(\ref{eq:wzwsu3}) actually differ by two contact terms
which transform in a nontrivial manner both under chiral rotations and under
the QCD renormalization group. In order to obtain a renormalization group
invariant anomaly, one must remove these contact terms \cite{KL1}. Since these two terms
involve the singlet axial vector field $\langle a_\mu \rangle$
and the derivative of the QCD vacuum angle, $\partial_\mu \theta$, they are not relevant
for the present work and can safely be neglected.

The first term in Eq.~(\ref{eq:wzwsu3}) contains the WZW term for the $SU(3)$ effective
theory
\beq  \label{eq:wzwlagr}
\int \; d^4 x \; {\cal L}_{\scriptscriptstyle{WZW}} (\bar{U}, v, \bar{a})
   \equiv S_{\scriptscriptstyle{WZW}} (\bar{U}, v, \bar{a}),
\eeq
while the second one is gauge invariant and does not contribute to the anomaly.
It is straightforward to show that the expression $B$
can be absorbed by contact terms of unnatural parity at fourth chiral order
\beqa \label{eq:unpar4}
d^4 x \;  \tilde{{\cal L}}_{p^4} &=&  i \tilde{L}_1 \, \bar{\psi} \, 
    \langle F_{r} D U D U^\dagger 
    + F_{l} D U^\dagger D U \rangle  +     2 \tilde{L}_2 \, \bar{\psi} \, 
\; \langle F_r U F_l U^\dagger \rangle \no \\
&&
+ 2 \tilde{L}_3 \, \bar{\psi} \,  
\; \langle F_r^2 + F_l^2 \rangle   +  i \tilde{L}_4  D \theta \langle  F_{r} D U U^\dagger  
      - F_{l} U^\dagger D U  \rangle \no \\
&&    
    +2 \tilde{L}_5 \, \bar{\psi} \, 
\; \Big( \langle F_r \rangle \langle F_r \rangle + 
   \langle F_l \rangle \langle F_l  \rangle \Big)
   + 2 \tilde{L}_6 \, \bar{\psi} \, 
\; \langle F_r \rangle \langle F_l  \rangle ,
\eeqa
where we employed the notation
\beq
F_{r} = dr -ir^2, \qquad F_l = dl -il^2 .
\eeq
%The potentials $V_i$ are gauge invariant functions of the variable $\bar{\psi}$.
%From parity it follows that the $V_i$ are odd functions of $\bar{\psi}$, except
%for $V_4$ which is even.
%
The vacuum angle $\theta$ has served its purpose and will be omitted for the rest of this section.
Since we are interested in radiative decays, we will furthermore set the external vector
and axial-vector fields
\beq
r = l = v = - e Q A 
\eeq
with $A$ being the photon field.
The anomalous Lagrangian ${\cal L}_{\scriptscriptstyle{WZW}} $ in Eq.~(\ref{eq:wzwlagr})
relevant for the two-photon decays at the one loop level reduces then to
\beqa \label{wzwvec}
S_{\scriptscriptstyle{WZW}} (\hat{U}, v)  &=& \int \; d^4 x \; 
              {\cal L}_{\scriptscriptstyle{WZW}} (\hat{U}, v)\no \\
  &=&  -  \frac{i N_c}{48 \pi^2} \int \; d^4 x \;
\langle \hat{\Sigma}_{\scriptscriptstyle{L}} \hat{U}^\dagger dv \hat{U} v 
           + \hat{\Sigma}_{\scriptscriptstyle{L}} \, v \, dv +
    \hat{\Sigma}_{\scriptscriptstyle{L}} \, dv \, v   
    - i \hat{\Sigma}^3_{\scriptscriptstyle{L}} \, v \rangle  - 
         (R \leftrightarrow L)  
\eeqa
with $U = e^{\frac{i}{3} \psi} \hat{U}$ and $\hat{\Sigma}_{\scriptscriptstyle{L}}= 
\hat{U}^\dagger d \hat{U}$.

The quark charge matrix $Q$ of the $u$- $d$- and $s$-quarks has usually been
assumed to be independent of the number of colors with $Q = \frac{1}{3} \mbox{diag} (2, -1,-1)$.
However, the cancellation of triangle anomalies requires $Q$ to depend on $N_c$ \cite{BW}
\beq
Q =  \mbox{diag} \Big( Q_u, Q_d, Q_s\Big)  
= \frac{1}{2}  \mbox{diag} \Big( \frac{1}{N_c} +1,\frac{1}{N_c} -1,\frac{1}{N_c} -1 \Big)
 =  \hat{Q} + \Big( 1 - \frac{N_c}{3}  \Big)  \frac{1}{2 N_c} \mathds{1} 
\eeq
with $\hat{Q} = \frac{1}{3} \mbox{diag} (2, -1,-1)$ being the conventional charge matrix,
while the second term is proportional to the baryon number and gives rise to the
Goldstone-Wilczek term.
The anomalous Lagrangian of Eq.~(\ref{wzwvec}) decomposes into the conventional
WZW Lagrangian of the $SU(3)$ theory with the charge matrix $\hat{Q}$ and a Goldstone-Wilczek term
which vanishes for $N_c =3$
\beq
S_{\scriptscriptstyle{WZW}} (\hat{U}, v) = S_{\scriptscriptstyle{WZW}} (\hat{U}, \hat{v})
+  \Big( 1 - \frac{N_c}{3}  \Big)  S_{\scriptscriptstyle{GW}} (\hat{U}, \hat{v})
\eeq
with $\hat{v} = -e \hat{Q} A$ and
\beqa
S_{\scriptscriptstyle{WZW}} (\hat{U}, \hat{v}) &=& \frac{ N_c e}{48 \pi^2} \int \;
   \big\langle \, (\hat{\Sigma}_{\scriptscriptstyle{L}}^3 
   - \hat{\Sigma}_{\scriptscriptstyle{R}}^3)\,  \hat{Q} \, \big\rangle \, A  \no \\
&-&
    \frac{i  N_c e^2}{48 \pi^2} \int \; 
   \big\langle \, 2 (\hat{\Sigma}_{\scriptscriptstyle{L}}
   - \hat{\Sigma}_{\scriptscriptstyle{R}})\,  \hat{Q}^2 
   + \hat{Q} \, (\hat{\Sigma}_{\scriptscriptstyle{L}} \hat{U}^\dagger \hat{Q} \hat{U}
-\hat{\Sigma}_{\scriptscriptstyle{R}} \hat{U} \hat{Q} \hat{U}^\dagger  ) \, \big\rangle \, dA \, A ,\\[0.3cm]
 S_{\scriptscriptstyle{GW}} (\hat{U}, \hat{v})  &=& \frac{ e}{48 \pi^2} \int \;
   \big\langle \, \hat{\Sigma}_{\scriptscriptstyle{L}}^3 \, \big\rangle \, A  
   - \frac{i e^2}{16 \pi^2} \int \; 
   \big\langle \, (\hat{\Sigma}_{\scriptscriptstyle{L}}
   - \hat{\Sigma}_{\scriptscriptstyle{R}})\,  \hat{Q} ) \, \big\rangle \, dA \, A .
\eeqa
It has been shown in \cite{BW} that the $N_c$ dependent part of the
Goldstone-Wilczek term cancels both the $\pi$-$ 2\gamma$ and the $\eta$-$ 2\gamma$ 
vertices of the WZW Lagrangian, yielding at tree level a decay width for
these decays which does not depend on $N_c$, if one neglects $\eta$-$\eta'$ mixing.
However, at the one-loop level other vertices involving kaons will contribute to the decays.
One can easily show that, {\it e.g.}, the vertex with two photons and $\pi^+, \pi^-, \pi^0$
of the WZW term is canceled by the $N_c$ dependent piece of the Goldstone-Wilczek
term, in agreement with the observation that the number of colors does not appear
in the effective theory for two flavors \cite{BW}.
The vertices involving kaons, on the other hand, do not cancel and an $N_c$ dependent
piece remains for the vertices. Consider as an example the vertex with
two photons and $\pi^0, K^+, K^-$ that contributes to the decay $\pi^0 \to \gamma \gamma$. 
The WZW term yields the vertex (neglecting mixing of the $\pi^0$ with the $\eta$-$\eta'$ system)
\beq
- \frac{ 5 N_c e^2}{72 \pi^2} K^+ K^- d\pi^0 dA A ,
\eeq
whereas the $N_c$ dependent piece of the Goldstone-Wilczek term leads to
\beq
 \frac{N_c e^2}{36 \pi^2} K^+ K^- d\pi^0 dA A .
\eeq
Clearly, both terms do not compensate and a dependence on $N_c$ remains in the final
expression for the vertex. It is therefore of interest to study the $N_c$ dependence
of the two-photon decays at the one-loop level.

%%%%%%%%%%%%%%%%%%%%%%%%%%%%%%%%%%%%%%%%%%%%%%%%%%%%%%%%%%%%%%%%%%%%%%%%%%%%%%
\section{Radiative decays at one-loop order}\label{sec:}

In the framework of large $N_c$ ChPT the expansion in powers of momenta and light
quark masses is combined
with the $1/N_c$ expansion by ordering the series according to
\beq
p = {\cal O}(\sqrt{\delta}) , \qquad m_q = {\cal O}(\delta), \qquad 1/N_c= {\cal O}(\delta).
\eeq
In this bookkeeping, the WZW term $S_{\scriptscriptstyle{WZW}}$  is of order
${\cal O}(\delta)$, whereas the one-loop diagrams of the decays 
involve the ratio $m_q/f^2$ with 
$f\sim {\cal O}(\sqrt{\delta})$ being the pseudoscalar decay constant in the chiral limit
and are thus of order ${\cal O}(\delta^3)$, {\it i.e.} next-to-next-to-leading order.

Our starting point is the WZW effective action of the $U(3)$ theory
\beqa
S_{\scriptscriptstyle{WZW}} (U, v) &=& \int \; d^4 x \; {\cal L}_{\scriptscriptstyle{WZW}} (U, v)\no \\
  &=&  -  \frac{i N_c}{48 \pi^2} \int \; d^4 x \;
\langle \Sigma_{\scriptscriptstyle{L}} U^\dagger dv U v  + \Sigma_{\scriptscriptstyle{L}} \, v \, dv 
      + \Sigma_{\scriptscriptstyle{L}} \, dv \, v   - i \Sigma^3_{\scriptscriptstyle{L}} \, v \rangle  - 
         (R \leftrightarrow L)  
\eeqa
We expand the quark charge matrix $Q$ in powers of  $1/N_c$
\beqa
Q &=& \frac{1}{2}  \mbox{diag} \Big( \frac{1}{N_c} +1,\frac{1}{N_c} -1,\frac{1}{N_c} -1 \Big)
 =  \frac{1}{2}  \mbox{diag} ( 1,-1,-1) + \frac{1}{2 N_c} \mathds{1} \no \\ % \mbox{diag} ( 1,1,1) \no \\
  &\equiv&   Q^{(0)}  + Q^{(1)} , 
\eeqa
where the superscript denotes the order in the combined large $N_c$ and chiral counting scheme,
{\it i.e.} $Q^{(0)}$ ($Q^{(1)}$) is of order ${\cal O}(1)$ (${\cal O}(\delta)$).
With $U = \exp (i  \phi )$,  one obtains for the three decays from 
$S_{\scriptscriptstyle{WZW}}$ the tree level contributions 
\beq \label{eq:lagrwzw}
d^4 x \; {\cal L}_{\scriptscriptstyle{WZW}} =  \frac{N_c e^2}{8 \pi^2} 
\langle d \phi\;   Q^2 \rangle \; A \; dA
=  - \frac{N_c e^2}{8 \pi^2} 
\big \langle \phi\;   [ (Q^{(0)})^2 + 2 Q^{(0)} Q^{(1)} + (Q^{(1)})^2 ] \big \rangle \; dA \; dA
\eeq
since $Q^{(0)}, Q^{(1)}$ commute with the diagonal entries of
$ \phi$---$ \phi^3, \phi^8, \phi^0$.
The terms on the right-hand side of Eq.~(\ref{eq:lagrwzw}) contribute
at orders $\delta,\delta^{2}$ and $\delta^{3}$, respectively, 
if one disregards the $N_c$-dependence of $\phi$ in which a factor $1/f = {\cal O}(1/\sqrt{N_c})$ 
has been absorbed.
Hence, within large $N_c$ ChPT the $\phi^3$ and $\phi^8$
decay amplitudes start at order ${\cal O}(\delta^{2})$, whereas the decay amplitude for
the singlet field $\phi^0$ is of order $\delta$.

At fourth chiral order the unnatural parity Lagrangian consists of more terms,
which are gauge invariant, see Eq.~(\ref{eq:unpar4}),
\beq
\tilde{{\cal L}}_{\scriptscriptstyle{eff}} =
{\cal L}_{\scriptscriptstyle{WZW}}+ \tilde{{\cal L}}_{p^4}
\eeq
with
\beqa
d^4 x \;  \tilde{{\cal L}}_{p^4} &=& 2 \tilde{V}_2 ( \bar{\psi} ) 
\; \langle F_r U F_l U^\dagger \rangle
+ 2 \tilde{V}_3 (\bar{\psi} ) 
\; \langle F_r^2 + F_l^2 \rangle  \no \\
&+&  2 \tilde{V}_5 ( \bar{\psi} ) 
\; \Big( \langle F_r \rangle \langle F_r \rangle + 
   \langle F_l \rangle \langle F_l  \rangle \Big)
+   
   2 \tilde{V}_6 ( \bar{\psi} ) 
\; \langle F_r \rangle \langle F_l  \rangle  + \ldots ,
\eeqa
where we have presented only the contact terms which contribute to the decays at the 
order we are working.

The potentials $\tilde{V}_i$ are odd functions in $\bar{\psi}$,
so that the leading contribution in the $1/N_c$ expansion is linear in $\bar{\psi}$.
The Lagrangian $\tilde{{\cal L}}_{p^4}$ can be expanded in powers of $1/N_c$
\beq
\tilde{{\cal L}}_{p^4}=  
 \tilde{{\cal L}}_{p^4}^{(2)} + \tilde{{\cal L}}_{p^4}^{(3)} + \ldots ,
\eeq
where the superscript denotes the order in the $\delta$ expansion
with 
\beq
d^4 x \;  \tilde{{\cal L}}_{p^4}^{(2)} = 2 \tilde{L}_2 \; \bar{\psi}
\; \langle F_r U F_l U^\dagger \rangle
+ 2 \tilde{L}_3 \; \bar{\psi} 
\; \langle F_r^2 + F_l^2 \rangle 
\eeq
and
\beq
d^4 x \;  \tilde{{\cal L}}_{p^4}^{(3)} = 
2 \tilde{L}_5 \; \bar{\psi} 
\; \Big( \langle F_r \rangle \langle F_r \rangle + 
   \langle F_l \rangle \langle F_l  \rangle \Big)
+   
   2 \tilde{L}_6 \; \bar{\psi} 
\; \langle F_r \rangle \langle F_l  \rangle .
\eeq
The contributions from $\tilde{{\cal L}}_{p^4}^{(2)}$ and 
$\tilde{{\cal L}}_{p^4}^{(3)}$ are of order ${\cal O}(p^4)$
and ${\cal O}(\frac{1}{N_c}p^4)$, respectively.
Setting $F_r = F_l = -e Q \; d A$, we obtain 
for $\tilde{{\cal L}}_{p^4}^{(2)}$
\beqa
d^4 x \;  \tilde{{\cal L}}_{p^4}^{(2)} &=& 2 e^2  [ \tilde{L}_2 + 2  \tilde{L}_3]
 \bar{\psi}
\big  \langle (Q^{(0)})^2 + 2 Q^{(0)} Q^{(1)} + (Q^{(1)})^2 \big \rangle \; dA \; dA \no \\
&=& e^2 k_1 
\Big( \frac{3}{4} - \frac{1}{2 N_c} \Big) \; \bar{\psi}\; dA \; dA 
\eeqa
where $k_1 = 2  ( \tilde{L}_2 + 2  \tilde{L}_3 ) $ and
the last term proportional to $(Q^{(1)})^2$ has been omitted,
since it is of order ${\cal O}(\delta^{4})$ and thus beyond our working precision. 
In a similar way, the terms from $\tilde{{\cal L}}_{p^4}^{(3)}$ reduce to 
\beq
d^4 x \;  \tilde{{\cal L}}_{p^4}^{(3)} = e^2 
k_2 \bar{\psi} \langle  Q^{(0)}  \rangle  \langle  Q^{(0)}  \rangle \; dA \; dA 
\eeq
with $k_2 = 2  ( 2 \tilde{L}_5 +  \tilde{L}_6 ) $. 

From the renormalization group invariance of the effective Lagrangian it follows 
that $k_1$ and $k_2$ transform as ({\it cf.} App. \ref{app:scal} for details)
\beqa
k_1^{ren} &=& Z_A k_1 - \frac{N_c (Z_A -1)}{24 \pi^2} \no \\
k_2^{ren} &=& Z_A k_2 ,
\eeqa
where $Z_A$ is the multiplicative renormalization constant of the singlet
axial current $A_\mu^0 = \frac{1}{2} \bar{q} \gamma_\mu \gamma_5 q$
which transforms as $(A_\mu^0)^{ren} = Z_A A_\mu^0$ under changes in the
QCD running scale.

At sixth chiral order the relevant terms for the decays read
\beqa
d^4 x \;  \tilde{{\cal L}}_{\chi} &=& i \tilde{W}_1 (\bar{\psi} )  \langle
 U \chi^\dagger F_r^2 + \chi^\dagger U F_l^2 \rangle 
+  i \tilde{W}_2 (\bar{\psi} )  \langle
\chi^\dagger F_r U F_l + U \chi^\dagger U F_l U^\dagger 
F_r \rangle \no \\
&+&
  i \tilde{W}_3 (\bar{\psi} )  \langle
 U \chi^\dagger\rangle \langle F_r^2 +  F_l^2 \rangle 
  + i \tilde{W}_4 (\bar{\psi} )  \langle
 U \chi^\dagger\rangle \langle U^\dagger F_r U F_l \rangle 
 \no \\
&+&
i \tilde{W}_5 (\bar{\psi} )  \langle F_r + F_l \rangle 
  \langle [ F_r U  + U F_l ] \chi^\dagger\rangle 
  \quad + \quad  h.c. \q .
\eeqa
The quark mass matrix $\mathcal{M} = \mbox{diag}(m_u,m_d,m_s)$
enters in the combination  $\chi  = 2 B \mathcal{M} $
with $B = - \langle  0 | \bar{q} q | 0\rangle/ \decay^2$ being the order
parameter of the spontaneous symmetry violation.
Expanding in powers of $1/N_c$ one obtains
\beq
\tilde{{\cal L}}_{\chi} = \tilde{{\cal L}}_{\chi}^{(2)} + \tilde{{\cal L}}_{\chi}^{(3)} 
\eeq
with the contributions $\tilde{{\cal L}}_{\chi}^{(2)}$ at ${\cal O}(N_c p^6)$
\beqa
d^4 x \;  \tilde{{\cal L}}_{\chi}^{(2)} &=& i \tilde{w}_1^{(0)}  \langle 
[ U \chi^\dagger - \chi  U^\dagger ] F_r^2
+ [ \chi^\dagger U  -  U^\dagger \chi ] F_l^2 \rangle  \no \\
&+&
i  \tilde{w}_2^{(0)} 
\langle [\chi^\dagger - U^\dagger \chi U^\dagger]
  F_r U F_l  +  [U \chi^\dagger U - \chi] 
  F_l U^\dagger F_r \rangle
\eeqa
and $\tilde{{\cal L}}_{\chi}^{(3)}$ at order ${\cal O}(p^6)$
\beqa
d^4 x \;  \tilde{{\cal L}}_{\chi}^{(3)} &=& 
 - \tilde{w}_1^{(1)}  \; \bar{\psi} \; \langle 
[ U \chi^\dagger + \chi  U^\dagger ] F_r^2
+ [ \chi^\dagger U  +  U^\dagger \chi ] F_l^2 \rangle  \no \\
&-&
  \tilde{w}_2^{(1)} \; \bar{\psi}  \; 
\langle [\chi^\dagger + U^\dagger \chi U^\dagger]
  F_r U F_l  +  [U \chi^\dagger U + \chi] 
  F_l U^\dagger F_r \rangle\no \\
&+&
i \tilde{w}_3^{(0)}  \langle 
 U \chi^\dagger - \chi  U^\dagger  \rangle  \langle F_r^2
+  F_l^2 \rangle  
+
i \tilde{w}_4^{(0)}  \langle  U \chi^\dagger - \chi  U^\dagger  \rangle
  \langle U^\dagger F_r U F_l \rangle \no \\
&+&
i \tilde{w}_5^{(0)} \langle F_r + F_l  \rangle
\langle F_r U \chi^\dagger - U^\dagger F_r \chi + U F_l \chi^\dagger
- F_l  U^\dagger \chi  \rangle ,
\eeqa
where the potentials $\tilde{W}_i$ have been expanded according to
$\tilde{W}_{i} = \tilde{w}_{i}^{(0)} + i \tilde{w}_{i}^{(1)} \bar{\psi} + {\cal O}(\bar{\psi}^2) $.

The explicitly symmetry breaking terms reduce to the structures 
\beq
d^4 x \;  \tilde{{\cal L}}_{\chi}^{(2)} = k_3 e^2 
\big  \langle  \phi \chi [(Q^{(0)})^2 + 2 Q^{(0)} Q^{(1)}] \big \rangle \; dA \; dA 
\eeq
with $k_3 =  - 4  (\tilde{w}_1^{(0)}+\tilde{w}_2^{(0)})$
and
\beq
d^4 x \;  \tilde{{\cal L}}_{\chi}^{(3)} = e^2  \Big(
k_4 \bar{\psi} \big  \langle  \chi (Q^{(0)})^2  \big \rangle
+k_5 \big  \langle  \phi \chi \big \rangle \big  \langle  (Q^{(0)})^2  \big \rangle 
+ k_6 \big  \langle Q^{(0)}\big \rangle \big  \langle  \phi \chi Q^{(0)}  \big \rangle
\Big) \; dA \; dA 
\eeq
with $k_4 =  - 4  (\tilde{w}_1^{(1)}+\tilde{w}_2^{(1)})$,
$k_5 =  - 2  (2 \tilde{w}_3^{(0)}+\tilde{w}_4^{(0)})$,
$k_6 =  - 8  \tilde{w}_5^{(0)}$,
respectively.
The scaling law for the parameter $k_4$ is given by ({\it cf.} App. \ref{app:scal})
\beq
k_4^{ren} = Z_A k_4 + \frac{1}{3} [Z_A -1] k_3 ,
\eeq
while the remaining parameters $k_3, k_5$ and $k_6$ remain put.

Having discussed the tree diagram contributions to the decays, we now turn to
the calculation of the loops at order $\delta^3$.
After expanding the WZW Lagrangian in the meson fields $\phi$ the contributing pieces 
at one-loop order read
\beqa \label{eq:loop}
d^4 x \;  {\cal L}_{\scriptscriptstyle{WZW}} &=&  - \frac{N_c e^2}{48 \pi^2 } 
\Big(  \langle d \phi [ \phi, [ \phi, Q^2]] \rangle -
 \langle d \phi [ \phi, Q] [ \phi, Q] \rangle  \Big) A \; dA \no \\
 && -  \frac{i  N_c e}{24 \pi^2 } \langle d \phi \; d \phi \; d \phi \; Q \rangle A + \ldots \q .
\eeqa
The mesons inside the loops do not undergo mixing, 
as $\phi^3,\phi^8$
and $\phi^0$ loops do not contribute.
The first two terms  in Eq.~(\ref{eq:loop})
contribute via tadpoles, whereas the last one represents a vertex of the unitarity 
correction, corresponding to Figure~1b. 
As the diagonal components of $\phi$ commute with $Q$, tadpoles with $\phi^3,\phi^8$
and $\phi^0$ do not contribute, 
and since the photon couples only to charged mesons, the unitarity corrections
arise due to charged meson loops.

\begin{figure}
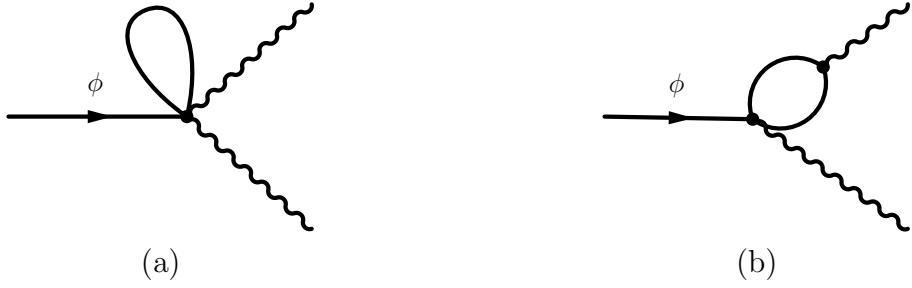

\centering
\begin{minipage}[b]{0.35\textwidth}
\centering
\includegraphics[scale=0.85]{fig.1a} \\
(a)
\end{minipage}
\hspace{0.1\textwidth}
\begin{minipage}[b]{0.35\textwidth}
\centering
\includegraphics[scale=0.85]{fig.1b} \\
(b)
\end{minipage}
\caption{One-loop diagrams contributing to $\phi \to \gamma \gamma$. 
         In (b) the crossed diagram is not shown.}
\label{fig1}
\end{figure}

At order ${\cal O}(\delta^{3})$ only the $Q^{(0)}$ piece in Eq.~(\ref{eq:loop}) contributes. From
$(Q^{(0)})^2= \frac{1}{4} \mathds{1}$ it follows then that the first term vanishes and
the tadpoles are entirely due to the second term.
Performing the loop integration, one obtains for the decays the tadpole contributions
\beq
\frac{ N_c e^2}{24 \pi^2 f^2}  \sum_{i=3,8,0} \Big(  d_\pi^i \Delta_\pi
+ d_K^i \Delta_K  \Big)  \phi^i \; dA \; dA
\eeq
with the tadpole given in dimensional regularization by
\beq
\Delta_\phi =   \int \frac{d^d l}{(2 \pi)^d} \frac{i}{l^2 - m_\phi^2 +i\epsilon} 
 = m_\phi^2 \Big[ 2 L +\frac{1}{16 \pi^2} \,  \ln \frac{m_\phi^2}{\mu^2} \Big] ,
\eeq
where $L$ contains the pole at $d=4$,
\beq
L = \frac{\mu^{d-4}}{16 \pi^2} \Big( \frac{1}{d-4} -\frac{1}{2} \big( \ln 4 \pi  +
\Gamma'(1) +1 \big)  \Big)
\eeq
and $\mu$ is the regularization scale.
The coefficients $d_\phi^i$ read
\beqa
d_\pi^3 &=&  \phantom{-} 0, \qquad \qquad  d_\pi^8= - \sqrt{\frac{2}{3}} ,
\qquad \qquad  d_\pi^0= - \frac{2}{\sqrt{3}} ; \no \\
d_K^3 &=& - \frac{1}{\sqrt{2}}, \qquad \qquad  d_K^8=  \frac{1}{\sqrt{6}} ,
\qquad \qquad  d_K^0= - \frac{2}{\sqrt{3}} .
\eeqa
Evaluating the unitarity corrections at order ${\cal O}(\delta^{3})$
for on-shell photons and replacing
$Q$ in Eq.~(\ref{eq:loop}) by $Q^{(0)}$, since the contribution from $Q^{(1)}$
is beyond the order we are working,
one obtains exactly the same contribution as for the tadpoles but with opposite sign.
Hence, the one-loop corrections to the decays at order ${\cal O}(\delta^{3})$
compensate each other and the first non-vanishing non-analytic piece will show up at order
${\cal O}(\delta^{4})$.
This is also in agreement with previous calculations in conventional ChPT, in which the chiral logarithms
were compensated completely by wavefunction renormalization and replacing $f$
by the physical decay constant $F_\phi$ in the tree level expression \cite{DHL, BBC1}. 
However, within large $N_c$ ChPT the $\phi^3$ and $\phi^8$
decay amplitudes start at order ${\cal O}(\delta^{2})$, so that the leading
non-analytic corrections to the physical decay constant and wavefunction renormalization
will contribute at order ${\cal O}(\delta^{4})$ and
do not affect the amplitude up to order ${\cal O}(\delta^{3})$.
Any divergences from the loop
diagrams discussed above could then only be renormalized by counter terms of the $p^6$ Lagrangian
of unnatural parity. This would be clearly in contradiction to previous results \cite{DW, BBC2}.

So far, the $\phi^8$ amplitude does not contain an explicit $N_c$ dependence due to the WZW
term, but both the $\phi^8$ and $\phi^0$ fields undergo mixing which results in the mass
eigenstates $\eta$ and $\eta'$. Here, we work in the isospin limit of equal up- and down-quark masses,
$\hat{m} = m_u = m_d$, so that the $\phi^3$ field decouples from the $\eta$-$\eta'$
system. In the next section, we will give an estimate on isospin breaking effects
due to different quark masses $m_u \ne m_d$.

In order to describe $\eta$-$\eta'$ mixing up to one-loop order, one must take into account
the following terms of the effective Lagrangian
\beq
{\cal L}_{\scriptscriptstyle{eff}} = {\cal L}^{(0)} + {\cal L}^{(1)} + {\cal L}^{(2)} + \ldots
\eeq
which reads at lowest order $\delta^0$
\beq
{\cal L}^{(0)} = \frac{f^2}{4} \langle D_\mu U ^\dagger D^\mu U \rangle 
+ \frac{f^2}{4} \langle \chi U ^\dagger +  U \chi^\dagger \rangle  - \frac{1}{2} \tau \bar{\psi}^2 . 
\eeq
The next-to-leading order Lagrangian ${\cal L}^{(1)} = {\cal O}(\delta)$ contains the terms 
\beqa
{\cal L}^{(1)} &=& L_5 \langle D_\mu U ^\dagger D^\mu U (\chi^\dagger U + U^\dagger \chi) \rangle 
+ L_8 \langle \chi^\dagger U \chi^\dagger U  + U^\dagger \chi U^\dagger \chi \rangle \no \\
&& + \frac{f^2}{12} \Lambda_1 D_\mu \psi D^\mu \psi + i \frac{f^2}{12} \Lambda_2 \bar{\psi} 
\langle \chi^\dagger U - U^\dagger \chi \rangle ,
\eeqa
and ${\cal L}^{(2)}$ is given by
\beqa
{\cal L}^{(2)} &=& L_4 \langle D_\mu U ^\dagger D^\mu U \rangle 
\langle \chi^\dagger U + U^\dagger \chi  \rangle
+ L_6 \langle \chi^\dagger U + U^\dagger \chi  \rangle^2
+ L_7 \langle \chi^\dagger U - U^\dagger \chi  \rangle^2 \no \\
&& + i L_{18} D_\mu \psi \langle D^\mu U^\dagger \chi - D^\mu U \chi^\dagger \rangle 
+ i L_{25} \bar{\psi} \langle U^\dagger \chi U^\dagger \chi - \chi^\dagger U \chi^\dagger U   \rangle 
+ {\cal O}(N_c p^6) .
\eeqa
The terms $\eta_0^2 \langle D_\mu U ^\dagger D^\mu U \rangle$ and 
$\eta_0^2 \langle \chi U ^\dagger +  U \chi^\dagger \rangle$ have been omitted in ${\cal L}^{(2)}$,
since the pertinent unknown coupling constants represent OZI violating corrections.
Moreover, as indicated in the last equation, counterterms  of order ${\cal O}(N_c p^6)$
with new unknown coupling constants will contribute at order $\delta^2$. We will neglect these contributions
throughout, assuming that they are of small size and do not alter our conclusions.

The fields $\phi^8$ and $\phi^0$ are related to the mass eigenstates $\eta$ and $\eta'$ via
\beqa
\phi^8 &=& \frac{\sqrt{2}}{F_\eta^8} \; [ \cos \vartheta^{(1)} - \sin \vartheta^{(0)} 
{\cal A}^{(1)} ] \;  \eta
+ \frac{\sqrt{2}}{F_\eta^8} \; [ \sin \vartheta^{(1)} + \cos \vartheta^{(0)} 
{\cal A}^{(1)} ] \;  \eta' \no \\
\phi^0 &=& \frac{\sqrt{2}}{\sqrt{3} F_{\eta'}^0} \; [ \cos \vartheta^{(1)}{\cal A}^{(2)}
 - \sin \vartheta^{(2)} {\cal B} ] \;  \eta 
+  \frac{\sqrt{2}}{\sqrt{3} F_{\eta'}^0} \; [ \sin \vartheta^{(1)}{\cal A}^{(2)}
 + \cos \vartheta^{(2)} {\cal B} ] \;  \eta' . 
\eeqa
The decay constants $F_\eta^8$ and $F_{\eta'}^0$ are defined by
\beqa  \label{decconst}
\langle 0 | \bar{q} \gamma_\mu \gamma_5  \lambda^8  q |\eta \rangle = i \sqrt{2} \; p_\mu F_\eta^8 \no \\
\langle 0 | \bar{q} \gamma_\mu \gamma_5  \lambda^0  q |\eta' \rangle = i \sqrt{2} \; p_\mu F_{\eta'}^0
\eeqa
with the normalization $\langle \lambda^a  \lambda^b  \rangle =  \delta^{ab}$,
while the angles $\vartheta^{(i)}$ correspond to the mixing angle up to order ${\cal O}(\delta^i)$
which arises in the diagonalization of the $\phi^8$-$\phi^0$ mass matrix
\beqa \label{angles}
\sin 2 \vartheta^{(0)} &=& - \frac{4 \sqrt{2}}{3} \frac{m_K^2 - m_\pi^2}{m_{\eta'}^2 - m_\eta^2} \no \\
\sin 2 \vartheta^{(1)} &=& \sin 2 \vartheta^{(0)} \Big( \frac{1 + \Lambda_2}{\sqrt{1 + \Lambda_1}} 
   + \frac{8}{F_\pi^2}  [2 L_8^{(r)} - L_5^{(r)} ] (m_K^2 - m_\pi^2) - 
   \frac{24}{F_\pi^4} L_5^{(r)} \tau  \Big)   \no \\
\sin 2 \vartheta^{(2)} &=&  \frac{2\sqrt{2}} {3[m_{\eta'}^2 - m_\eta^2]} \Bigg(     
   2 [\mcirc{\pi} - \mcirc{K} ] \;\frac{1 + \Lambda_2}{\sqrt{1 + \Lambda_1}} 
   + \frac{32}{F_\pi^2} \; [m_\pi^2 -m_K^2] \; m_K^2 \; \frac{2 L_8^{(r)} - 
                                 3 L_{25}}{\sqrt{1 + \Lambda_1}}\no \\ 
&&
 +   \frac{16}{F_\pi^2} \; [m_\pi^2 -m_K^2] \; [ 2 m_K^2 + m_\pi^2] \; (2 L_6^{(r)} + 2 L_7-L_4^{(r)} ) 
  \no \\ 
&&
- \frac{24}{F_\pi^4} \; [m_\pi^2 -m_K^2] \; ( 2 L_5^{(r)} 
         + 3 L_{18}^{(r)}) \; \tau \; (1 - \frac{5}{4} \Lambda_1)
+  \frac{64}{F_\pi^6} \; [m_\pi^2 -m_K^2] \; [ 11 m_K^2 + m_\pi^2] \; L_5^{(r) \; 2} \tau  \no \\ 
&& 
-  \frac{16}{F_\pi^2} \; [m_\pi^2 -m_K^2] \; [ 7 m_K^2 + m_\pi^2] \; L_5^{(r)}  
    \; (1 - \frac{1}{4} \Lambda_1) \no \\ 
&& 
 + \frac{8}{F_\pi^2} \; [m_\pi^2 -m_K^2] \; [ 5 m_K^2 + m_\pi^2] \;
   ( 2 L_5^{(r)} (1 - \frac{1}{3} \Lambda_2) - L_{18}^{(r)})   
- \frac{3}{ 2 F_\pi^2} \; m_\pi^2 \Delta_\pi^{(r)}  + \frac{1}{  F_\pi^2} \; m_K^2 \Delta_K^{(r)}  \no \\ 
&& 
- \frac{1}{ 3 F_\pi^2} \; \Delta_\eta^{(r)}  \big( m_\pi^2 [\sfrac{5}{2} \cos^2 \vartheta^{(0)}  
+ \sfrac{1}{\sqrt{2}} \sin 2 \vartheta^{(0)}  + 2 \sin^2 \vartheta^{(0)}  ]\no \\ 
&& \qquad \qquad \; \;
 - m_K^2 [4 \cos^2 \vartheta^{(0)}  
+ 2 \sqrt{2} \sin 2 \vartheta^{(0)}  + 2 \sin^2 \vartheta^{(0)}  ])  \no \\ 
&& 
- \frac{1}{ 3 F_\pi^2} \; \Delta_{\eta'}^{(r)}  \big( m_\pi^2 [\sfrac{5}{2} \sin^2 \vartheta^{(0)}  
- \sfrac{1}{\sqrt{2}} \sin 2 \vartheta^{(0)}  + 2 \cos^2 \vartheta^{(0)}  ]\no \\ 
&& \qquad \qquad \; \;
 - m_K^2 [4 \sin^2 \vartheta^{(0)}  
- 2 \sqrt{2} \sin 2 \vartheta^{(0)}  + 2 \cos^2 \vartheta^{(0)}  ])
 + {\cal O}(N_c p^6) \Bigg)
\eeqa
where $ F_\pi \approx 93$ MeV is the pion decay constant defined in a 
similar way as in Eq.~(\ref{decconst}), and $m_\eta, m_{\eta'}$ are the diagonal
entries of the $\eta$-$\eta'$ mass matrix.
Furthermore, $L_i^{(r)}$ and $\Delta_\phi^{(r)} = m_\phi^2/(16 \pi^2) \ln (m_\phi^2/\mu^2) $ 
are the finite parts of the LECs 
and the loops, respectively, after renormalization. It is straightforward to verify that the angles
$\vartheta^{(i)}$ do not depend on the regularization scale of the effective theory.

The quantities ${\overset{\circ}{m}\vphantom{m}_{\pi}}$ and ${\overset{\circ}{m}\vphantom{m}_{K}}$
denote the pion and kaon masses at leading order
\beqa
\mcirc{\pi} &=& 2 B \hat{m} =m_\pi^2 \Big( 1 - \frac{8}{ F_\pi^2} m_\pi^2 [2 L_8^{(r)} - L_{5}^{(r)}] 
 - \frac{8}{ F_\pi^2} (2 m_K^2 + m_\pi^2) [2 L_6^{(r)} - L_{4}^{(r)}] - \frac{1}{2 F_\pi^2} 
   \Delta_\pi^{(r)} \no \\
&& \qquad +  \frac{1}{6 F_\pi^2} [\cos^2 \vartheta^{(0)} + 2 \sin^2 \vartheta^{(0)}] \Delta_\eta^{(r)} 
+ \frac{1}{6 F_\pi^2} [\sin^2 \vartheta^{(0)} + 2 \cos^2 \vartheta^{(0)}] \Delta_{\eta'}^{(r)} \Big) , \no \\
\mcirc{K} &=& B ( \hat{m} + m_s ) = m_K^2 \Big( 1 - \frac{8}{ F_\pi^2} m_K^2 [2 L_8^{(r)} - L_{5}^{(r)}] 
 - \frac{8}{ F_\pi^2} (2 m_K^2 + m_\pi^2) [2 L_6^{(r)} - L_{4}^{(r)}]  \no \\
&& \qquad -  \frac{1}{3 F_\pi^2} \cos 2 \vartheta^{(0)}  \Delta_\eta^{(r)} \Big)
+  \frac{1}{12 F_\pi^2} [4 m_K^2 \cos^2 \vartheta^{(0)} -
   ( 3 m_{\eta'}^2 + m_\pi^2 ) \sin^2 \vartheta^{(0)}] \Delta_{\eta'}^{(r)}  .
\eeqa
The expressions ${\cal A}^{(i)}$ and ${\cal B}$ read
\beqa \label{eq:zetmix}
{\cal A}^{(1)} &=& \frac{8 \sqrt{2}}{3 F_\pi^2} L_5^{(r)} [m_K^2 - m_\pi^2]  \no \\ 
{\cal A}^{(2)} &=& \frac{4 \sqrt{2}}{3 F_\pi^2} [m_K^2 - m_\pi^2] \bigg(  
   \frac{2 L_5^{(r)} + 3 L_{18}^{(r)}}{(1 + \Lambda_1)^{1/4}} +  \frac{8}{3 F_\pi^2}
   L_5^{(r) \; 2} [- m_K^2 + 13 m_\pi^2] - \frac{32}{ F_\pi^2} L_5^{(r)} L_8^{(r)} [m_K^2 + m_\pi^2] 
\bigg)\no \\ 
{\cal B} &=& 1 + \frac{4}{3 F_\pi^2} [2 m_K^2 + m_\pi^2] \bigg( 3 L_4^{(r)} - L_5^{(r)}
  + \frac{2 L_5^{(r)} + 3 L_{18}^{(r)}}{\sqrt{1 + \Lambda_1}}
  -  \frac{3 L_4^{(r)} + L_5^{(r)} + 3 L_{18}^{(r)}}{1 + \Lambda_1} \bigg) \no \\ 
  && + \frac{64}{9 F_\pi^4} L_5^{(r) \; 2} \; [3 m_K^4 - 4 m_\pi^2 m_K^2 + 3 m_\pi^4] . 
\eeqa
For the $\phi^8$ decay we have only kept the pieces up to next-to-leading order,
since the terms beyond that order contribute at ${\cal O}(\delta^4)$.
In the case of the $\phi^0$ decay, on the other hand, one must keep also the contributions
at next-to-next-to-leading order.
The values of the couplings $\Lambda_1, \Lambda_2, L_{18}^{(r)}$ and $L_{25}$ are not known
and depend on the running scale of QCD. As they represent OZI violating corrections,
we will omit them, but, strictly speaking, we cannot expect that all neglected terms vanish
at the same scale.
Furthermore, the parameter $\tau$ is related to the mass of the $\eta'$ in the chiral limit which was
estimated in \cite{etap} to be about 850 MeV. This translates into a value of
$\tau \approx 1 \times 10^{-3}$ GeV$^4$.

Including the mixing from Eq.~(\ref{angles}) we obtain the amplitudes
\beqa \label{eq:ampl}
&& e^2 \bigg[   \frac{B_{\pi}}{F_\pi} \pi^0 + \Big( \frac{B_{\eta}}{F^8_\eta} 
 [ \cos \vartheta^{(1)} - \sin \vartheta^{(0)}  {\cal A}^{(1)} ] 
 + \frac{B_{\eta'}}{F^0_{\eta'}} [ \cos \vartheta^{(1)}{\cal A}^{(2)}
 - \sin \vartheta^{(2)} {\cal B} ] \Big) \eta \no \\
&& \qquad +  \Big( \frac{B_{\eta}}{F^8_\eta} [ \sin \vartheta^{(1)} + \cos \vartheta^{(0)} 
{\cal A}^{(1)} ]   +   \frac{B_{\eta'}}{F^0_{\eta'}} [ \sin \vartheta^{(1)}{\cal A}^{(2)}
 + \cos \vartheta^{(2)} {\cal B} ] \Big) \eta' \bigg] dA \; dA ,
\eeqa
where $\pi^0$ is related to the $\phi^3$ field via
\beq
\pi^0 = \frac{f}{\sqrt{ 2 Z_{\pi}} } \phi^3 = \frac{F_\pi}{\sqrt{ 2} } \phi^3
\eeq
with the pion $Z$-factor 
\beq
\sqrt{Z_\pi} = 1 - \frac{4}{f^2} m_\pi^2 L_5^{(r)}
\eeq
and the decay constant
\beq
F_\pi = f \Big(1 + \frac{4}{f^2} m_\pi^2 L_5^{(r)} \Big)
\eeq
to the order we are working.
The coefficients $B_\phi$ read
\beqa 
B_{\pi} &=&   - \frac{N_c }{4 \sqrt{2} \pi^2 } \langle \lambda_3 Q^2 \rangle
+ \sqrt{2} k_3  \big  \langle  \lambda_3 
\chi [(Q^{(0)})^2 + 2 Q^{(0)} Q^{(1)}] \big \rangle \no \\
&& + \sqrt{2} k_5 \big  \langle \lambda_3 \chi \big \rangle 
\big  \langle  (Q^{(0)})^2  \big \rangle 
+ \sqrt{2} k_6 \big  \langle Q^{(0)}\big \rangle \big  \langle  
\lambda_3 \chi Q^{(0)}  \big \rangle    \no \\[0.3cm]
B_{\eta} &=&   - \frac{N_c }{4 \sqrt{2} \pi^2 } \langle \lambda_8 Q^2 \rangle
+ \sqrt{2} k_3  \big  \langle  \lambda_8 
\chi [(Q^{(0)})^2 + 2 Q^{(0)} Q^{(1)}] \big \rangle \no \\
&& + \sqrt{2} k_5 \big  \langle \lambda_8 \chi \big \rangle 
\big  \langle  (Q^{(0)})^2  \big \rangle 
+ \sqrt{2} k_6 \big  \langle Q^{(0)}\big \rangle \big  \langle  
\lambda_8 \chi Q^{(0)}  \big \rangle    \no \\[0.3cm]
B_{\eta'} &=& - \frac{N_c }{4 \sqrt{6} \pi^2 } \langle Q^2 \rangle
+ \sqrt{6} k_1 \big  \langle (Q^{(0)})^2 + 2 Q^{(0)} Q^{(1)}  \big \rangle
+ \sqrt{6} k_2 \langle  Q^{(0)}  \rangle  \langle  Q^{(0)}\rangle \no \\
&& + \sqrt{\frac{2}{3}} k_3  \big  \langle  
\chi [(Q^{(0)})^2 + 2 Q^{(0)} Q^{(1)}] \big \rangle 
+ \sqrt{6} k_4  \big  \langle  \chi (Q^{(0)})^2 \big \rangle 
\no \\
&& + \sqrt{\frac{2}{3}} k_5 \big  \langle \chi \big \rangle 
\big  \langle  (Q^{(0)})^2  \big \rangle 
+ \sqrt{\frac{2}{3}} k_6 \big  \langle Q^{(0)}\big \rangle \big  \langle  \chi Q^{(0)}  \big \rangle   .
\eeqa
Due to the $N_c$ dependence of the quark charge matrix $Q$
the expressions $B_{\pi}$ and $B_{\eta}$ start at order $\delta^2$, whereas
$B_{\eta'}$ contains a piece of order ${\cal O}(\delta)$.
Substituting these relations into Eq.~(\ref{eq:ampl}) yields the decay widths
\beqa   \label{ampl}
\Gamma_{\pi^0 \to \gamma \gamma} &=& \alpha^2 \pi m_{\pi^0}^3 \; 
 \Big |\,\frac{B_{\pi}}{F_\pi} \, \Big |^{\,2} , \no \\
\Gamma_{\eta \to \gamma \gamma} &=& \alpha^2 \pi m_{\eta}^3 \; \; 
 \Big |\, \frac{B_{\eta}}{F^8_\eta} 
 [ \cos \vartheta^{(1)} - \sin \vartheta^{(0)}  {\cal A}^{(1)} ] 
 + \frac{B_{\eta'}}{F^0_{\eta'}} [ \cos \vartheta^{(1)}{\cal A}^{(2)}
 - \sin \vartheta^{(2)} {\cal B} ]  \, \Big |^{\,2} , \no \\
\Gamma_{\eta' \to \gamma \gamma} &=& \alpha^2 \pi m_{\eta'}^3 \; \; 
 \Big |\, \frac{B_{\eta}}{F^8_\eta} [ \sin \vartheta^{(1)} + \cos \vartheta^{(0)} 
{\cal A}^{(1)} ]   +   \frac{B_{\eta'}}{F^0_{\eta'}} [ \sin \vartheta^{(1)}{\cal A}^{(2)}
 + \cos \vartheta^{(2)} {\cal B} ]  \, \Big |^{\,2} 
\eeqa
with $\alpha= e^2/4 \pi$.
In the $\delta$ expansion the leading order contribution to the decay width of the
$\eta$ is given due to mixing by the leading contribution in $B_{\eta'}$,
and is comparable in size with the $ B_{\eta}$ portion.
The numerical values will be discussed in detail in the next section.

%%%%%%%%%%%%%%%%%%%%%%%%%%%%%%%%%%%%%%%%%%%%%%%%%%%%%%%%%%%%%%%%%%%%%%%%%%%%%%%
\section{Numerical analysis} \label{sec:num}

Equations~(\ref{angles}) and (\ref{eq:zetmix}) are utilized to obtain values for the mixing angles
$\vartheta^{(i)}$ and the expressions ${\cal A}^{(i)}, {\cal B} $, respectively.
For the LECs $L_{4,5,6,8}^{(r)}, L_7$, we take values which follow from matching
the $U(3)$ theory to the $SU(3)$ framework by integrating out the singlet field
\cite{KL1, HS2} at the regularization scale $\mu=1$ GeV,
$L_{4}^{(r)} (\mu) = -0.5, L_{5}^{(r)} (\mu) = 1.0, L_{6}^{(r)} (\mu) = -0.3, 
L_{7} = -0.3, L_{8}^{(r)} (\mu) = 0.7$ 
(all in units of 10$^{-3}$).
Note that integrating out the singlet field only alters the LECs $L_{6}^{(r)}, L_{7}$
and $L_{8}^{(r)}$ and their values in the $U(3)$ framework are within the 
phenomenologically determined error ranges of the $SU(3)$ LECs.
Moderate variations of these LECs yield small changes in the decay amplitudes
with the largest changes induced by variations in $L_7$ roughly at the 10\% level.
Our conclusions are therefore not altered, if slightly different values for the LECs
are employed.  
Using the experimental values for the pseudoscalar meson masses,
we obtain
$\vartheta^{(0)} = -21.8^\circ, \vartheta^{(1)} = -15.8^\circ, \vartheta^{(2)} = -19.8^\circ$.

By employing Eq.~(\ref{ampl}), we can now fit the ratios 
$B_{\pi}/F_\pi, \,B_{\eta}/F^8_\eta, \, B_{\eta'}/F^0_{\eta'}$ to the decay widths
$\Gamma_{\pi^0 \to \gamma \gamma},$  
$\Gamma_{\eta \to \gamma \gamma},$  
$\Gamma_{\eta' \to \gamma \gamma}$.
The experimental values for the decay widths are \cite{PDG}
\beqa
\Gamma_{\pi^0 \to \gamma \gamma} &=&  7.74 \pm 0.55 \; \mbox{eV} , \no \\
\Gamma_{\eta \to \gamma \gamma} &=&  0.465 \pm 0.045 \; \mbox{keV} , \no \\
\Gamma_{\eta' \to \gamma \gamma} &=&  4.28 \pm 0.34 \; \mbox{keV} ,
\eeqa
and the fit to the central values yields
\beqa
B_{\pi}/F_\pi &=&  -0.133  \; \mbox{GeV}^{-1}, \no \\
B_{\eta}/F^8_\eta &=&  -0.0522  \; \mbox{GeV}^{-1}, \no \\
B_{\eta'}/F^0_{\eta'} &=& -0.192  \; \mbox{GeV}^{-1} .
\eeqa
For the pion decay constant $F_\pi$ we employ the physical value $F_\pi \approx 93$ MeV,
while $F^8_\eta$ can be extracted from a one-loop calculation 
with $F^8_\eta = 1.34 F_\pi$ \cite{KL2}.
It is consistent to take the one-loop results for $F_\pi$ and $F^8_\eta$, since the
difference with respect to the next-to-leading order expressions
shows up at $\delta^4$ in the decay amplitude and is, therefore, beyond our working precision.
The values for $B_{\pi}/F_\pi$ and $B_{\eta}/F^8_\eta$ from the fit
are close to the contributions from the anomalous WZW term, 
$B_\pi^{\scriptscriptstyle{WZW}}/F_\pi = -0.136$ GeV$^{-1}$,
$B_\eta^{\scriptscriptstyle{WZW}}/F^8_\eta = -0.0587$ GeV$^{-1}$,
indicating that the portions from the counter terms of unnatural parity are
small. Omitting higher orders beyond $\delta^3$, they contribute with
a relative strength of about $ 2\%$ to $B_\pi$ and $ 10\%$ to $B_\eta$.
In order to get an estimate for $F^0_{\eta'}$, it is thus justified to assume
that the counter term combination in $B_{\eta'}$ is small as well.
Of course, both the counter terms and $F^0_{\eta'}$ depend on the renormalization scale
$\mu_{\scriptscriptstyle{QCD}}$, but we will assume that for a certain range of
$\mu_{\scriptscriptstyle{QCD}}$ the counter term contributions are negligible.
For such $\mu_{\scriptscriptstyle{QCD}}$, the ratio 
$B_{\eta'}^{\scriptscriptstyle{WZW}}/F^0_{\eta'}$ is then reproduced by setting
$F^0_{\eta'} \approx 1.16 \; F_\pi$, a value slightly larger than in previous calculations \cite{DHL, BBC1}.

In particular, we would like to investigate, whether a clear statement can be given
on the number of colors by utilizing the $1/N_c$ expansions of the decay amplitudes.
The cancellation of Witten's global $SU(2)_L$ anomaly requires $N_c$ to be odd
\cite{Wit}. The standard model with $N_c=1$ is without strong interactions.
We will therefore restrict ourselves to a comparison
of the numerical results for $N_c=3$ and $N_c=5$.
Setting all non-anomalous contact terms of unnatural parity to zero, the decay width 
for the $\eta$ in a world
with $N_c=5$ reads $\Gamma_{\eta \to \gamma \gamma}^{N_c =5}= 1.002  \; \mbox{keV} $
to be compared with the decay width in the real world with three colors,
$\Gamma_{\eta \to \gamma \gamma}^{N_c =3}= 0.511  \; \mbox{keV} $.
For the $\eta'$ we obtain $\Gamma_{\eta' \to \gamma \gamma}^{N_c =3}= 4.21  \; \mbox{keV} $ and
$\Gamma_{\eta' \to \gamma \gamma}^{N_c =5}= 12.8  \; \mbox{keV} $.
The experimental values for the $\eta$ and $\eta'$ decays clearly rule out $N_c=5$
and varying the values for the omitted coupling constants of the counter terms
within realistic ranges does not alter this conclusion.

Finally, we would like to give an estimate on the $N_c$ dependence of the $\pi^0$
decay width due to different up- and down-quark masses. In the case of different 
up- and down-quark masses, the $\phi^3$ field undergoes mixing with both the
$\phi^8$ and $\phi^0$ field.
In order to get an estimate, we will restrict ourselves to the mixing at leading order
in the $\delta$ expansion. The fields $\phi^3, \phi^8$ and $\phi^0$
are then related to the mass eigenstates via
\beqa
\phi^3 &=& \frac{\sqrt{2}}{F_\pi} \Big( \pi^0 - \epsilon \eta - \epsilon' \eta' \Big)  \no \\
\phi^8 &=& \frac{\sqrt{2}}{F^8_\eta} \Big( \cos \vartheta^{(0)} \; ( \eta + \epsilon \pi^0) 
        + \sin \vartheta^{(0)} \; ( \eta' + \epsilon' \pi^0) \Big) \no \\
\phi^0 &=& \frac{\sqrt{2}}{\sqrt{3} F^0_{\eta'}} \Big( - \sin  \vartheta^{(0)} \; ( \eta + \epsilon \pi^0) 
        + \cos \vartheta^{(0)} \; ( \eta' + \epsilon' \pi^0) \Big)
\eeqa
with the mixing parameters 
\beqa \label{mixpar}
\epsilon_0 &=& \frac{\sqrt{3}}{4}   \frac{m_d -m_u}{m_s - \hat{m}} , \no \\
\epsilon &=& \epsilon_0 \frac{\cos \vartheta^{(0)} 
               - \sqrt{2} \sin  \vartheta ^{(0)}}{1 + \frac{1}{\sqrt{2}}\tan \vartheta^{(0)} } , \no \\
\epsilon' &=& \epsilon_0 \frac{\sin \vartheta^{(0)} 
               + \sqrt{2} \cos  \vartheta^{(0)} }{1 - \frac{1}{\sqrt{2}} \mbox{ctg} \; \vartheta^{(0)} }.    
\eeqa
The parameter $\epsilon_0$ can be expressed in terms of physical meson masses
by applying Dashen's theorem \cite{Dashen:1969eg}, which implies the identity of the pion
and kaon electromagnetic mass shifts up to ${\cal O}(e^2p^2)$
\begin{equation} \label{dash}
\epsilon_0 = \frac{m^2_{K^0}-m^2_{K^\pm}+m^2_{\pi^\pm}-m^2_{\pi^0}}{\sqrt{3} (m_\eta^2 - m_\pi^2)} .
\end{equation}
There have been estimates in the literature that Dashen's theorem is significantly 
violated at higher orders due to chiral symmetry breaking effects 
\cite{Donoghue:1992ha, Donoghue:1993hj, Bijnens:1993ae}.
On the other hand, a recent non-perturbative approach to the hadronic decays
of $\eta$ and $\eta'$ indicated that higher order corrections to this low-energy theorem
may be small \cite{BB1}.
In any case, the isospin-violating effects in the decay widths constitute a small correction,
so that it is safe to employ Dashen's theorem in the present work.
%, leading to
%$\epsilon_0 \approx 0.01 $, $\epsilon \approx 0.02 $ and $\epsilon' \approx 0.003$.

This time the fit to the central experimental values yields 
\beqa
B_{\pi}/F_\pi &=&  -0.134 \; \mbox{GeV}^{-1} ,\no \\
B_{\eta}/F^8_\eta &=&  -0.0598  \; \mbox{GeV}^{-1}, \no \\
B_{\eta'}/F^0_{\eta'} &=& -0.208  \; \mbox{GeV}^{-1} .
\eeqa
and setting $F^0_{\eta'} = 1.07 F_\pi$ in the
ratio $B_{\eta'}^{\scriptscriptstyle{WZW}}/F^0_{\eta'}$ reproduces the fitted value.

It should be emphasized that we fitted our results to the current world
average value for $\Gamma_{\pi^0 \to \gamma \gamma}= 7.74 \pm 0.55 \; \mbox{eV} $ \cite{PDG}.
On the other hand, it is possible to estimate in a model-dependent way
the size of the counter term contributions.
In \cite{GBH}, {\it e.g.}, the values of the counter term contributions to the 
$\pi^0$ decay have been estimated
by means of a QCD sum rule for the general three-point function
involving the pseudoscalar density and two vector currents. Within that approach
the authors find a slightly enhanced width of
$\Gamma_{\pi^0 \to \gamma \gamma}= 8.10 \pm 0.08 \; \mbox{eV} $.

A comparison of the numerical results for $N_c =3$ and $N_c =5$ and setting all
non-anomalous contact terms of unnatural parity to zero yields
$\Gamma_{\pi^0 \to \gamma \gamma}^{N_c =3}= 8.00  \; \mbox{eV}$,
$\Gamma_{\pi^0 \to \gamma \gamma}^{N_c =5}= 8.18  \; \mbox{eV}$ for the pion decay,
$\Gamma_{\eta' \to \gamma \gamma}^{N_c =3}= 0.456 \; \mbox{keV} $,
$\Gamma_{\eta' \to \gamma \gamma}^{N_c =5}= 0.920  \; \mbox{keV} $
for the $\eta$ and
$\Gamma_{\eta' \to \gamma \gamma}^{N_c =3}= 13.7  \; \mbox{keV} $,
$\Gamma_{\eta' \to \gamma \gamma}^{N_c =5}= 4.28  \; \mbox{keV} $
for the $\eta'$.
Again, $N_c=5$ is ruled out by the $\eta$ and $\eta'$ decays,
but no rigorous statement can be made for the $\pi^0$, although the value for
$N_c=3$ is in better agreement with the current world average.

%%%%%%%%%%%%%%%%%%%%%%%%%%%%%%%%%%%%%%%%%%%%%%%%%%%%%%%%%%%%%%%%%%%%%%%%%%%%%%%
\section{Conclusions} \label{sec:concl}

In the present work, we investigated the two-photon decays of $\pi^0, \eta$
and $\eta'$ in the combined $1/N_c$ and chiral expansions. The cancellation of 
triangle anomalies in the standard model requires the quark charges to depend
on $N_c$. We have shown that the WZW term of the $U(3)$ effective theory decomposes
into the conventional anomalous $SU(3)$ WZW Lagrangian, a Goldstone-Wilczek
term and counter terms of unnatural parity which involve the singlet field $\eta_0$.
The independence of the $\pi^0$ and $\eta$ decay amplitudes on $N_c$
which was shown in \cite{BW} to occur at tree-level due to partial cancellations of the WZW
term with a Goldstone-Wilczek term, persists at one-loop order, although the
vertices of the pertinent loop graphs {\it do} exhibit an $N_c$ dependence.

We performed a one-loop calculation including counter terms up to next-to-next-to
leading order in large $N_c$ ChPT in which $\eta$-$\eta'$ mixing has also been
taken into account up to one-loop order.
Within the bookkeeping of large $N_c$ ChPT, the leading contribution to the
$\eta$ decay arises from mixing with the $\eta'$.

From a fit to the experimental decay widths and under the assumption that
higher orders beyond our working precision can be neglected it follows that contributions 
from the counter terms are small.
Since the cancellation of Witten's global $SU(2)_L$ anomaly requires $N_c$ to be odd
and a world with $N_c=1$ has no strong interactions, we compare the cases
$N_c=3$ and $N_c=5$. 
The numerical results of the $\eta$ and $\eta'$ decay widths for $N_c=3$ 
are close to the experimental values
and clearly rule out the case $N_c=5$. 
We have furthermore given an estimate on the $N_c$ dependence of the pion decay
due to different up- and down-quark masses by taking $\pi^0$-$\eta$-$\eta'$ mixing
at leading order into account. 
The $N_c$ dependence of the $\pi^0$ decay is smaller than the experimental
uncertainty and is therefore not suited to extract the number of colors.
We conclude that both the $\eta$ and the $\eta'$ decay show clear evidence that we live in a
world with three colors.

It has been pointed out in \cite{BW} that at tree level the process
$\eta \to \pi^+ \pi^- \gamma$ is proportional to $N_c^2$ and should replace the
textbook process $\pi^0 \to \gamma \gamma$ lending support to $N_c=3$.
It will be thus of interest to investigate the decays $\eta \to \pi^+ \pi^- \gamma$ 
and $\eta' \to \pi^+ \pi^- \gamma$ within the framework of large $N_c$ ChPT \cite{BL}.

%%%%%%%%%%%%%%%%%%%%%%%%%%%%%%%%%%%%%%%%%%%%%%%%%%%%%%%%%%%%%%%%%%%%%%%%%%%%%%%
\section*{Acknowledgements}

The author is grateful to Edisher Lipartia and
Robin Ni{\ss}ler for useful discussions.

%%%%%%%%%%%%%%%%%%%%%%%%%%%%%%%%%%%%%%%%%%%%%%%%%%%%%%%%%%%%%%%%%%%%%%%%%%%%%%
\appendix 

%%%%%%%%%%%%%%%%%%%%%%%%%%%%%%%%%%%%%%%%%%%%%%%%%%%%%%%%%%%%%%%%%%%%%%%%%%%%%%%
\section{Scaling behavior of the coupling constants} \label{app:scal}
In this appendix, we derive the scaling behavior of the coupling constants
which contribute to the decays.
The transformation properties of the constants $\tilde{L}_{2,3}$ have already
been given in \cite{KL1} and we merely quote the result here.
The quantities $\tilde{L}_{2}$ and $\tilde{L}_{3}$ are renormalized according to
\beq
\tilde{L}_{2}^{ren} = Z_A \tilde{L}_{2} - \kappa , \qquad
\tilde{L}_{3}^{ren} = Z_A \tilde{L}_{3} - \kappa 
\eeq
with
\beq
\kappa = \frac{N_c (Z_A -1)}{144 \pi^2} .
\eeq
Since the singlet field $\bar{\psi}$ scales as $\bar{\psi}^{ren} = Z_A^{-1} \bar{\psi}$,
the scaling behavior of $\tilde{L}_{2,3}^{ren}$ ensures that
the Lagrangian ${\cal L}_{\scriptscriptstyle{WZW}}+\tilde{{\cal L}}_{p^4}^{(2)}$ 
remains invariant to order $\delta^2$ under changes of the QCD running scale.
In order to study the transformation properties of $\tilde{L}_{5,6}$
we rewrite $ \tilde{{\cal L}}_{p^4}^{(3)}$
\beqa
\tilde{{\cal L}}_{p^4}^{(3)} &=& 
2 \tilde{L}_5 \; \bar{\psi} 
\; \Big( \langle F_r \rangle \langle F_r \rangle + 
   \langle F_l \rangle \langle F_l  \rangle \Big)
+   
   2 \tilde{L}_6 \; \bar{\psi}  
\; \langle F_r \rangle \langle F_l  \rangle  \no \\
&=&
2 \bar{\psi}  \; [2 \tilde{L}_5 + \tilde{L}_6 ] 
\; \langle dv \rangle \langle dv\rangle + 
2 \bar{\psi}  \; [2 \tilde{L}_5 - \tilde{L}_6 ] 
   \langle da \rangle \langle da \rangle 
\eeqa
This yields the transformation properties
\beqa
(2\tilde{L}_{5}+\tilde{L}_{6})^{ren} &=& Z_A (2\tilde{L}_{5}+\tilde{L}_{6}) , \no \\
(2\tilde{L}_{5}-\tilde{L}_{6})^{ren} &=& Z_A^3 (2\tilde{L}_{5}-\tilde{L}_{6})
- \frac{N_c (Z_A^3 -1)}{432 \pi^2} ,
\eeqa
so that the Lagrangian ${\cal L}_{\scriptscriptstyle{WZW}}+\tilde{{\cal L}}_{p^4}^{(2)}+
\tilde{{\cal L}}_{p^4}^{(3)}$ remains renormalization group invariant.

On the other hand, the contact terms $\tilde{W}_1$ and $\tilde{W}_2$ in the Lagrangian 
of sixth chiral order can be written as
\beqa \label{eq:scalep6}
i \tilde{W}_1 (\bar{\psi} ) e^{\frac{i}{3}\bar{\psi}}  \langle
 \bar{U} \chi^\dagger F_{\bar{r}}^2  + \chi^\dagger \bar{U} F_{\bar{l}}^2 \rangle 
+ i \tilde{W}_2 (\bar{\psi} ) e^{\frac{i}{3}\bar{\psi}}  \langle
 \chi^\dagger F_{\bar{r}}  \bar{U} F_{\bar{l}} +  \bar{U} \chi^\dagger  \bar{U} 
  F_{\bar{l}} \bar{U}^\dagger F_{\bar{r}} \rangle +  h.c. + \ldots  . &&
\eeqa
The ellipsis in Eq.~(\ref{eq:scalep6}) denotes terms with more than one flavor trace
which involve contributions from other contact terms and are irrelevant for the discussion
of the scaling behavior of  $\tilde{W}_1$ and $\tilde{W}_2$.
From (\ref{eq:scalep6}) we obtain the transformation properties
\beqa
\tilde{W}_1 (x)^{ren} &=& \tilde{W}_1(Z_A x) e^{\frac{i}{3} (Z_A-1)x} , \no \\
\tilde{W}_2 (x)^{ren} &=& \tilde{W}_2(Z_A x) e^{\frac{i}{3} (Z_A-1)x} .
\eeqa
Expanding the potentials $\tilde{W}_i$ in the singlet field $\bar{\psi}$ yields
for the two leading expansion coefficients
\beqa
(\tilde{w}_i^{(0)})^{ren} &=& \tilde{w}_i^{(0)} , \no \\
(\tilde{w}_i^{(1)})^{ren} &=& Z_A \tilde{w}_i^{(1)} + \frac{1}{3} (Z_A -1) \tilde{w}_i^{(0)} , 
\qquad i=1,2 .
\eeqa
%

%%%%%%%%%%%%%%%%%%%%%%%%%%%%%%%%%%%%%%%%%%%%%%%%%%%%%%%%%%%%%%%%%%%%%%%%%%%%%%

\end{document}